\documentclass[article]{aa}  
\usepackage{natbib}
\bibpunct{(}{)}{;}{a}{,}{,}
\usepackage{multirow}
\usepackage[dvips]{graphicx}
\usepackage{txfonts}
\usepackage{rotating}
\usepackage{setspace}
\usepackage{threeparttable}

\begin{document}
\newcommand{\kms}{km~s$^{-1}$}

   \title{CO observations of water-maser post-AGB stars and detection of a high-velocity outflow in IRAS $15452-5459$}
\titlerunning{CO in water-maser post-AGB stars}

   \author{L. Cerrigone
          \inst{1}
          \and
          K. M. Menten \inst{2} \and T. Kami{\'n}ski \inst{2} \fnmsep%
          }

\institute{Centro de Astrobiolog{\'\i}a, INTA-CSIC, ctra de Ajalvir km 4, 28850 Torrej{\'o}n de Ardoz, Spain \\
\email{cerrigone@cab.inta-csic.es}
\and
   Max-Planck-Institut f\"ur Radioastronomie (MPIfR), Auf dem H\"ugel 69, D-53121 Bonn \\
             }

   \date{}

 
  \abstract
   {}
   {Many aspects of the evolutionary phase in which Asymptotic Giant Branch stars (AGB stars) are in transition to become Planetary Nebulae (PNe) are still poorly understood. An important question is how the circumstellar  envelopes of AGB stars switch from spherical symmetry to the axially symmetric structures frequently observed in PNe. In many cases there is clear evidence that the shaping of the circumstellar envelopes of PNe is linked to the formation of jets/collimated winds and their interaction with the remnant AGB envelope. Because of the short evolutionary time, objects in this phase are rare, but their identification provides valuable probes for testing evolutionary models. }%
   {We have observed (sub)millimeter CO rotational transitions with the APEX telescope in a small sample of stars hosting high-velocity OH and water masers. These targets  are supposed to have recently left the AGB, as indicated by the presence of winds traced by masers, with velocities larger than observed during that phase. We have carried out observations of several CO lines, ranging from $\mathrm{J}=2-1$ up to $\mathrm{J}=7-6$. }
   {In IRAS $15452-5459$ we detect a fast molecular outflow in the central region of the nebula and estimate a mass-loss rate between $1.2\times10^{-4}$~M$_\odot$~yr$^{-1}$ (assuming optically thin emission) and $4.9\times10^{-4}$~M$_\odot$~yr$^{-1}$ (optically thick emission). {We model the SED of this target taking advantage of our continuum measurement at 345 GHz to constrain the emission at long wavelengths. {For a distance of 2.5 kpc, we estimate a luminosity of 8000~L$_\odot$ and a dust mass of $\sim$0.01~M$_\odot$}. Through the flux in the $[$CII$]$ line (158~$\mu$m), we calculate a total mass of about 12 M$_\odot$ for the circumstellar envelope, but the line is likely affected by interstellar contamination.}}
   {}


   \maketitle
%

\section{Introduction}
When a low- or intermediate-mass star has completed its core helium burning, it goes through the AGB phase, characterized by heavy mass loss. Dust formation is very efficient in AGB outflows
and about 1--2\% of the mass of the envelope is condensed into dust grains. After leaving the AGB, the mass-loss rate
drops substantially and the star may become optically visible, as its dusty shell disperses. Eventually, once it
reaches a temperature of 20,000--30,000 K, the central star will start to ionize its circumstellar envelope (CSE)
and a Planetary Nebula will form \citep{habing}.
Dusty CSEs re-radiate absorbed stellar light and show clear signatures in the far-infrared (IR) spectrum. In particular, a \lq\lq color-color\rq\rq~diagram introduced by \citet{vanandhabing} has allowed for a classification of sources from the ratios of flux densities measured with the InfraRed Astronomical Satellite (IRAS) . 

The heavy mass-loss of AGB stars (with rates up to several 10$^{-4}$ M$_\odot$ yr$^{-1}$) is one of the most important pathways for mass
return from stars to the Interstellar Medium (ISM). 

After the AGB phase, the star \lq\lq somehow\rq\rq~develops a less massive (10$^{-7}$ M$_\odot$~yr$^{-1}$) but faster (up to 10$^{-3}$ km s$^{-1}$) wind that
sweeps up the circumstellar material, possibly creating shocks and high-density shells. UV photons due to the
increase in stellar temperature can excite, ionize and dissociate molecules, heat the dust, ionize the gas (when T$_\mathrm{eff} > 20$--$30\times10^{3}$ K).


In the CSEs of many O-rich AGB stars, maser emission is detected from SiO, H$_2$O and OH molecules. 
The conditions for the formation of maser emission from these different molecules are typically met in a progression of zones. The SiO masers arise from the closest regions to the central star, then H$_2$O and OH at progressively farther distances. 
In stars with thicker envelopes, longer periods and higher mass-loss rates - so-called OH/IR stars - the spectral profiles of H$_2$O and OH masers are often double-peaked, and the velocity range covered by the OH (10--25 km~s$^{-1}$) is usually larger than that of the H$_2$O feature \citep{engels}. When the mass-loss process ends (i.e., at the end of the AGB) and the circumstellar envelope is progressively diluted, the maser lines {may} start to disappear. {In a spherically symmetric CSE}, first SiO lines vanish, then H$_2$O, and finally OH (Lewis, 1998). 
Given these considerations,  OH/IR stars with detached circumstellar shells, whose presence can be inferred from doubly or multiply peaked OH-maser spectra with a large velocity spread, are good candidate objects for studying the post-AGB phase.

The members of a small but growing sub-class of water-maser stars  have been dubbed \lq\lq \textit{water-fountain}\rq\rq~nebulae and show unique spectral
and spatial characteristics \citep{likkel}. The velocity spread of their H$_2$O spectra is very large, up to several 100~km~s$^{-1}$, and frequently larger than that of their OH spectra. High-resolution interferometric observations have shown that their masers are in collimated jets  with dynamical ages shorter than 150 yr \citep{bobo}. Water-maser post-AGB stars and particularly water fountains represent a very short evolutionary phase right after the end of the AGB; therefore they provide a unique opportunity for studying the physical conditions in the molecular envelopes right when the shaping agent is starting its action.

{One such star} is IRAS~$15452-5459$, which shows SiO, H$_2$O and OH masers.
It was first selected by \citet{vanderveen} within a sample of sources detected with IRAS as a candidate star in transition from the AGB
to the PN phase. They required that their targets fall within regions IIIb, IV, and V of the van der Veen \& Habing IRAS
color-color diagram (i.e., evolutionary sequence from the AGB to the PN), and thus be bright at 12~$\mu$m ($>2$~Jy) and have 
a small IRAS variability index ($<50$). IRAS~$15452-5459$ (hereafter I15452) falls within region IV of that diagram, where variable stars with very thick O-rich circumstellar shells are expected. The authors assumed a stellar temperature of 10$^4$~K and a luminosity of 
{$1.5\times10^4$~L$_\odot$}, estimating a distance of about 2.5~kpc.

\citet{oudmaijer} performed IR K-band spectral observations and found CO ro-vibrational lines in absorption, which made them conclude
that the central star should have a temperature in the range of 3000--5000 K. 


This lower stellar temperature actually appears more consistent with the fact that I15452's CSE hosts masers in all the species found in CSEs. The object was included in the survey of OH masers carried out by \citet{telintel91}, who detected a peculiar 
OH maser at 1612 MHz with 4 peaks. The features span roughly $-20$ to $-100$ km~s$^{-1}$ and their peaks are found around $-83$, $-67$, $-47$, 
and $-33$ km~s$^{-1}$, with a central velocity around $-57$~km~s$^{-1}$. Subsequent observations with the Australia Telescope
Compact Array confirmed the detection of the components at larger velocities and their association with the star \citep{deacon04}.

\citet{deacon07} detected the H$_2$O maser line at 22 GHz. They found an irregular spectrum with many emission features, covering roughly $-30$ to $-80$~km~s$^{-1}$ and pointing to 
a complex masing region. 
They carried out three runs of observations over  one year and concluded that the spectral features are stable in velocity 
over a timescale of a few months, but the peak intensity of the individual features and the integrated spectral flux density vary
 dramatically ($\sim$70\%). They also report the detection of an SiO maser at 86 GHz with an unusual doubly-peaked profile,
 with features centered around $-40$ and $-75$ km~s$^{-1}$. The detection of OH, H$_2$O, and SiO masers and the near-IR CO absorption indicate that the source is an O-rich star
that has very recently evolved off the AGB and has started its transition across the HR diagram.

\section{Observations and results}
\subsection{IRAS $15452-5459$}
We observed I15452 with the APEX\footnote{This publication is based on data acquired with the Atacama Pathfinder Experiment (APEX). APEX is a collaboration between the Max-Planck-Institut f{\"u}r Radioastronomie, the European Southern Observatory, and the Onsala Space Observatory.} 12 m telescope, which is located at an altitude of 5100 m on Llano de Chajnantor, in Chile \citep{apex}. Observations were performed in both continuum and spectral-line mode, to investigate its CSE dusty and  molecular components.

\subsubsection{CO observations}
The CO observations were carried out on  2009 October 16 and 19. The receivers APEX-2 and FLASH \citep{flash1,flash} were used to observe the CO $\mathrm{J}=3-2$ (345.796 GHz) and $\mathrm{J}=4-3$ (461.041 GHz) rotational lines, respectively. The Fast Fourier Transform Spectrometer (FFTS) backend available at APEX allows for a spectral coverage of about 1.8 GHz, split up in 8192 channels, which gives velocity resolutions of 0.21 and 0.12 km~s$^{-1}$ at 345 and 460 GHz, respectively \citep{ffts}.

Since the source is known from infrared imaging to be extended over about $25''$, we pointed the telescope at three different positions. One pointing was performed at the central coordinates of the source (right ascension 15:49:11.38; declination -55:08:51.6), one  was off-set by $+7''$ in both RA and DEC (this is indicated in the following as the North-Eastern pointing) and a third pointing was si\-mi\-larly off-set by $-7''$ (indicated in the following as the South-Western pointing); each off-set pointing was therefore about $10''$ away from the central coordinates. 

 The observations were carried out using a position-switching mode, with the reference position located $1^\circ$ away in azimuth. 
The data reduction consisted in averaging together the scans performed for each  pointing and removing a low-order ($\leq2$) polynomial baseline. To enhance the signal-to-noise ratio, the data were smoothed to resolutions of 0.63~km~s$^{-1}$ for the $\mathrm{J}=3-2$ line and 2.4 km~s$^{-1}$ for the $\mathrm{J}=4-3$ line. 
The raw spectra are stored in the $T_{\rm A}^{\star}$ scale and can be converted to main-beam brightness temperature using $T_{{\rm mb}}$ = $T_{\rm A}^{\star}\frac{F_{eff}}{B_{eff}}$, where  $T_{\rm A}^{\star}$ is the antenna temperature corrected for atmospheric attenuation using the chopper-wheel method, $F_{eff}$ is the telescope forward efficiency (for APEX $F_{eff}\sim$0.95 at 345 GHz and 0.97 at 460 GHz) and  $B_{eff}$ is the main-beam efficiency, estimated to be 0.73 at 345 GHz and 0.61 at 460 GHz \citep{apex}. The uncertainty of the absolute intensity scale is estimated to be about $\pm$20\%. Regular pointing checks were made on strong circumstellar CO sources and typically found to be consistent within $\sim$3$\hbox{$^{\prime\prime}$ }$. The FWHM of the main beam, $\theta_{{\rm mb}}$, is 17.3$\hbox{$^{\prime\prime}$ }$ at 345 GHz and 13.3$\hbox{$^{\prime\prime}$ }$ at 460 GHz. Table~\ref{tab:obs_summary} summarizes the integration times, system temperatures and rms noise for each pointing.

\begin{table*}\centering
 \begin{tabular}{lccccccc}
\hline\hline\noalign{\smallskip}
Pointing & \multicolumn{3}{c}{$\mathrm{J}=3-2$} & &\multicolumn{3}{c}{$\mathrm{J}=4-3$}  \\
             & Time & T$_{sys}$  & RMS & & Time  &  T$_{sys}$ & RMS \\
             & min  & K          & mK  & &min    & K         & mK \\
\hline\noalign{\smallskip}
North-East$^{\mathrm{a}}$  & 2   & 444  & 125    & & 8  & 847  & 89  \\
Center  & 3.3 & 460  & 87    &  & 8  & 776  & 83  \\
South-West$^{\mathrm{b}}$  & 0.7 & 444  & 213   &  & 8  & 798  & 80  \\
\hline       
\end{tabular}
\caption{Summary of the observations of IRAS~$15452-5459$ {performed on 2009-10-16 ($\mathrm{J}=3-2$) and 2009-10-19 ($\mathrm{J}=4-3$)}. The rms noise is calculated after smoothing the spectra to 0.63  and 2.4 km~s${-1}$ for the $\mathrm{J}=3-2$ and $\mathrm{J}=4-3$ line, respectively.}
\label{tab:obs_summary}
\begin{list}{}{}
\item[$^{\mathrm{a}}$] {The North-East pointing was offset by 7$''$ in RA and 7$''$ in DEC.}
\item[$^{\mathrm{b}}$] {The South-West pointing was offset by $-7''$ in RA and $-7''$ in DEC.}
\end{list}
\end{table*}

 \subsubsection{Continuum observations}
On 2007 May 28, we observed I15452 with the Large APEX BOlometer CAmera (LABOCA) at 870~$\mu$m (345~GHz). LABOCA is an array of 295 bolometers, operated in total-power mode at a temperature of $\sim 285$~mK. The noise-weighted mean point-source sensitivity of the array determined from on-sky integrations is 55~mJy~s$^{1/2}$  per channel \citep{siringo}.

The absolute calibration is achieved by observations of pla\-nets, namely Mars, Uranus, and Neptune.  
The overall calibration accuracy for LABOCA is about 10\% \citep{siringo}. 

A major issue in performing sub-mm observations is the atmospheric opacity. Two independent methods are typically used for determining the sky opacity at APEX. The first one relies on the measurement of precipitable water vapor performed every minute by the APEX radiometer along the line of sight \citep{pardo}. This approach is limited by the knowledge of the passband, the applicability of the atmospheric model and the accuracy of the radiometer. The second method uses \textit{skydips}, i.e. measurements of the power of the atmospheric emission as a function of the airmass. The values of sky opacity  resulting from skydip analysis are robust, though up to $\sim $30\% larger than those obtained by the radiometer. {This systematic difference is not well understood, therefore,} to reconcile the results obtained with the two methods, a linear combination of radiometer and skydip values has been used, so as to obtain the most consistent calibrator fluxes at all elevations{, as indicated by \citet{siringo}.}

The observations were carried out in raster-spiral mapping mode. In the basic spiral mode, spiral scans are performed at a constant angular speed and an increasing radius around the pointing position, producing fully sampled maps of the whole field of view. These spirals are the preferred observing mode for sources brighter than a few Jy. For fainter sources, the basic spiral pattern can be combined with a raster mapping mode (raster-spirals) on a grid of pointing positions resulting in a denser sampling of the maps and longer integration time \citep{siringo}.

The data reduction was performed with the Bolometer array data Analysis (BoA) software, specifically developed for APEX bolometer data.
BoA allows the observer to perform all of the necessary reduction steps: opacity correction, flux calibration, bad channel flagging, removal of correlated noise, despiking, data weighting, creation of a map, and image analysis \citep{siringo}.

The nebula shows a flux density at 345 GHz of $260\pm40$ mJy, where the error includes the rms noise and the absolute calibration uncertainty.

\subsection{CO in water-maser stars}
Using the APEX telescope, we made an attempt to find CO rotational emission in {six} selected objects known to host high-velocity  water masers, like I15452. The sample included four water-fountain nebulae toward which no CO emission has previously been reported. All targets are listed in Table~\ref{obslog}. The six sources were observed in the CO($2-1$) transitions with the APEX-1 receiver, five of those were observed also in the CO($3-2$) transitions with the APEX-2 receiver, and two of them were also observed with the CHAMP+ array \citep{champ} giving spectra centered on the CO($6-5$) and ($7-6$) lines. The observing dates and noise levels are given in Table~\ref{obslog}. The telescope beam widths and main beam efficiencies at the observed frequencies are given in Table~\ref{apex}. As a back-end, we used the digital FFTS, which provided spectra at resolutions between 0.2--0.6~\kms\ and bandwidths of at least several hundred \kms.

The stars were  selected from  currently-available maser surveys (\citet{deacon04} and \citet{deacon07}) as having doubly peaked maser profiles with velocities larger than typical AGB values (10--25 km~s$^{-1}$), IRAS colors typical of post-AGB stars, and being observable from the APEX site. All of them host H$_2$O masers and are supposed to have just approached their post-AGB evolution. 
\citet{imai} observed 9 of the 13 water fountains currently  known  in the CO $\mathrm{J}=3-2$ line with the Atacama Submillimeter Telescope Experiment (ASTE). Their rms noise ranged from 4 to 62 mK and they detected the line in 2 targets.
In our selection, we have  excluded  the water fountains observed by \citet{imai} and included the newly-discovered water fountain IRAS~$19190+1102$ \citep{day}. This leaves us with {4 water fountains and an additional 2 stars with water masers.}

\begin{figure}\centering
\includegraphics[width=0.45\textwidth]{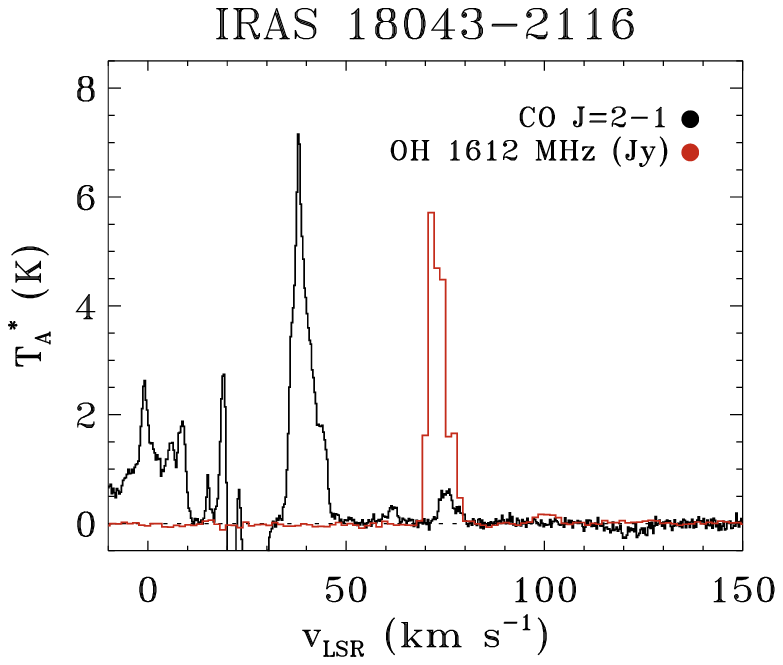}
 \includegraphics[width=0.45\textwidth]{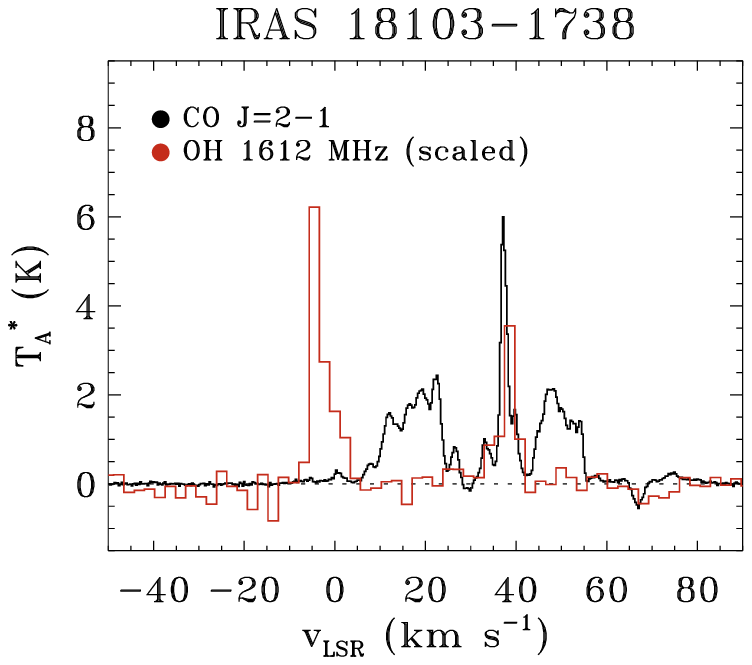}
\includegraphics[width=0.45\textwidth]{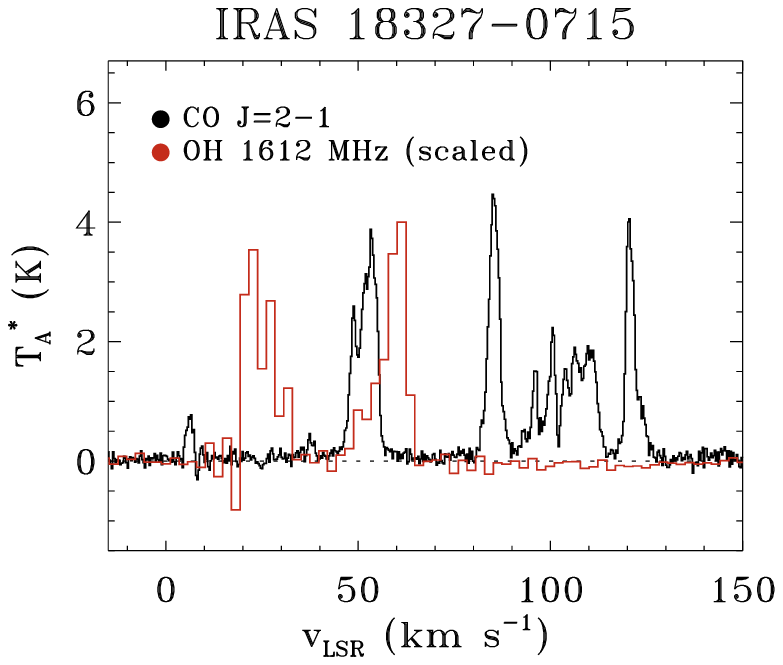}
\caption{Overlay of OH maser features on our CO spectra for targets with maser data available through on-line archives. The maser data are from \citet{sevenster} and they are available through VizieR \citep[http://vizier.u-trasbg.fr/viz-bin/VizieR]{vizier}.} 
\label{cooh_overlays}
\end{figure}

The observations in the CO($2-1$) and CO($3-2$) lines were obtained in position switching (total power) mode. Many of the sources that we observed are located close to the Galactic plane, therefore their reference positions were chosen 1--3\degr\ away from the Galactic plane, to avoid emission at the OFF positions. All the observed lines of sight cross interstellar molecular clouds, which give rise to prominent emission features, especially strong in the CO($2-1$) line. For {all sources other than} OH~$12.8-0.9$, these interstellar emission features occur in the range of velocities where the maser emission is observed. This strongly hampers the identification of any potentially existing CO emission from the circumstellar material in our objects. Indeed, for a majority of the objects, strong interstellar lines overlap in spectral regions close to the expected source velocity. For IRAS~$18103-1738$, we find one feature that is centered around the stellar velocity, when we compare the CO and OH maser spectra (Fig.~\ref{cooh_overlays}), but the CO feature may be of interstellar origin. In the case of two other objects, IRAS~$18043-2116$ and IRAS~$18327-0715$, we found relatively weak emission features that appear close to the expected central velocities, i.e. at 38 \kms\ and 87 \kms\ for IRAS~$18043-2116$ and IRAS~$18327-0715$, respectively (Fig.~\ref{cooh_overlays}). These emission features are candidates for the circumstellar emission but their interstellar origin cannot be excluded {without careful imaging observations and accurate positions}. For these two sources, we obtained deep integrations in the CO($6-5$) line (simultaneously with a spectrum in the CO($7-6$) line). In these high rotational transitions the contamination from interstellar clouds of low excitation should be negligible and the circumstellar gas can be identified more easily, if excited to high $J$ levels. In both cases the spectra do not show any emission in CO($6-5$) nor in ($7-6$) at the noise levels given in Table~\ref{obslog}. Our search for CO emission in these sources is therefore inconclusive. 

\begin{figure}\centering
 \includegraphics[width=0.45\textwidth]{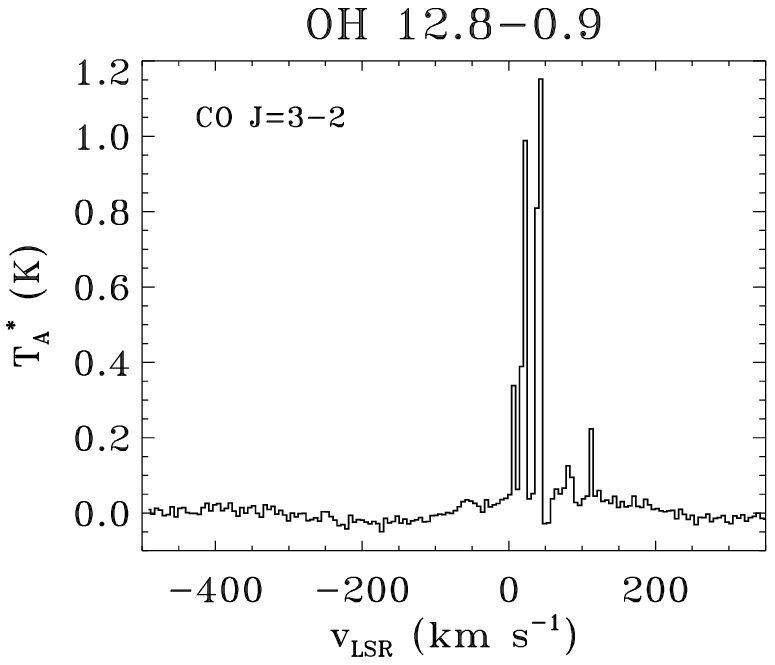}
 \includegraphics[width=0.45\textwidth]{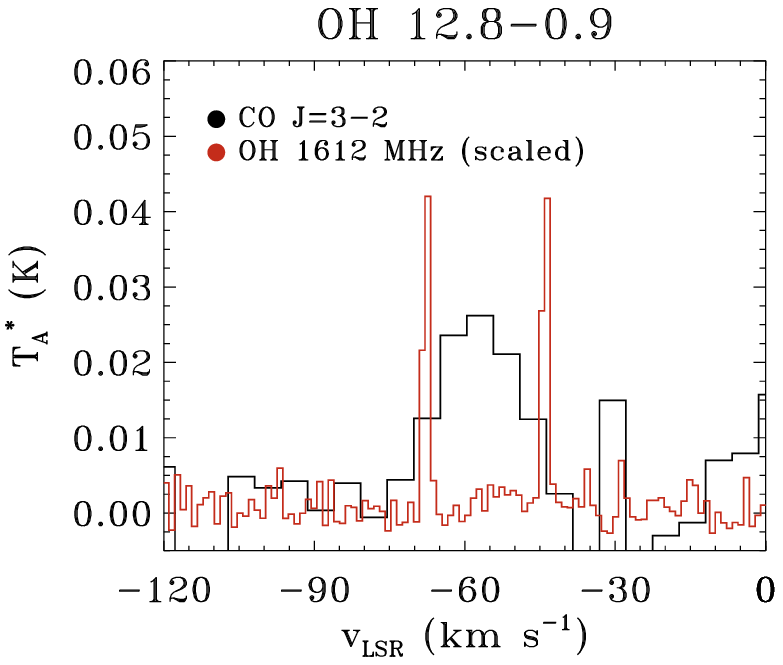}
\caption{CO spectrum observed toward OH~$12.8-0.9$. In the bottom, the overlay of the smoothed spectrum and the OH maser spectrum is shown, as in Fig.~\ref{cooh_overlays}.}\label{figOH12}
\end{figure}

In the case of OH~$12.8-0.9$, the maser features appear in a velocity range that is not contaminated by strong interstellar CO emission. We detected weak and relatively broad emission features in CO($2-1$) and ($3-2$) lines covering  the same velocity range as bracketed by the maser peaks of H$_2$O and OH, i.e. between $-70$ and $-35$ \kms\ \citep{bobo}. The detection is marginal, but the feature clearly appears above the 3$\sigma$ noise level after smoothing the spectra to a resolution of 5--6~\kms, as can be seen in Figure~\ref{figOH12}. The fact that it appears in both lines and the good match with the maser velocities support the detection, although more sensitive observations are desirable.   

\begin{table*}\centering 
\caption{Characteristics of CO observations in water fountain sources.}
\label{obslog}
\begin{tabular}{cc cc cc cc}
\hline\hline\noalign{\smallskip}
IRAS ID & OH &\multicolumn{2}{c}{Coordinates}& $V_{\rm LSR}$\tablefootmark{a}& 
Date  & CO  & rms\tablefootmark{b}\\
&& $\alpha_{\rm J2000}$ & $\delta_{\rm J2000}$ &(km~s$^{-1}$)&of observation&transition&mK\\
\hline\noalign{\smallskip}
$18327-0715$& $024.692+00.235$\tablefootmark{c}& 18:35:29.82& --07:13:09.4 & ~~42.0 & 2010-07-06  & $2-1$ &77 \\    
                &       &            & 	            &        & 2010-07-07  & $3-2$ &19 \\
                &       &            & 	            &        & 2010-07-09  & $6-5$ &48 \\
                &       &            & 	            &        & 2010-07-09  & $7-6$ &165 \\
$18043-2116$& $009.097-00.392$\tablefootmark{c}& 18:07:21.20& --21:16:14.0 & ~~87.3 &2010-07-05,06& $2-1$ &79 \\    
                &       &            & 	            &        & 2010-07-07  & $3-2$ &26 \\
                &       &            & 	            &        & 2010-07-09  & $6-5$ &91 \\
                &       &            & 	            &        & 2010-07-09  & $7-6$ &298 \\
$15103-5754$& $320.9063-00.2929$\tablefootmark{d}& 15:14:18.47& --58:05:20.9 & --47.0 & 2010-07-06  & $2-1$ &71 \\   
                &       &            & 	            &        & 2010-07-07  & $3-2$ &22 \\
$19190+1102$ & $046.2559-01.4764$\tablefootmark{d}& 19:21:25.30&  +11:08:40.0 & ~~50.0 & 2010-07-06  & $2-1$ &53 \\    
                &       &            & 	            &        & 2010-07-07  & $3-2$ &25 \\
--         & $012.816-00.895 $\tablefootmark{c} & 18:16:49.23& --18:15:01.8 & --55.8 &2010-07-05,06& $2-1$ &25 \\   
                &       &            & 	            &        & 2010-07-07  & $3-2$ &23 \\
$18103-1738$& $012.973+00.133$\tablefootmark{c} & 18:13:20.24& --17:37:17.3 & ~~17.0 & 2010-11-16  & $2-1$ &35 \\                
\hline\end{tabular}
\tablefoot{
\tablefoottext{a}{Object LSR radial velocity from \citet{deacon04} and \citet{deacon07}.}
\tablefoottext{b}{Noise rms level of the spectrum given per bin of 1~km~s$^{-1}$ and in the $T_{\rm mb}$ units.}
\tablefoottext{c}{From \citet{sevenster}}
\tablefoottext{d}{From SIMBAD\footnote{http://simbad.u-strasbg.fr/simbad/}}
}
\end{table*}

\begin{table}\centering 
\caption{Telescope specifications for the observed frequencies.}
\label{apex}
\begin{tabular}{cc cc cc cc}
\hline\hline\noalign{\smallskip}
CO         & Frequency&Beam\tablefootmark{a}&$\eta_{\rm mb}$\tablefootmark{b}  \\   
transition & (GHz)    &(arcsec)             &                 \\
\hline\noalign{\smallskip}
$2-1$ & 230.538 & 27   & 0.75\\
$3-2$ & 345.796 & 17.3 & 0.73\\
$4-3$ & 461.041 & 13.3 & 0.61\\
$6-5$ & 691.473 & 9    & 0.56\\
$7-6$ & 806.652 & 7.7  & 0.43\\
\hline\end{tabular}
\tablefoot{
\tablefoottext{a}{Half-power beam width.}
\tablefoottext{b}{Antenna main beam efficiency.}}
\end{table}

\section{IRAS~$15452-5459$: CO spectra}
\begin{figure*}
\centering
\includegraphics[height=0.93\textheight,clip]{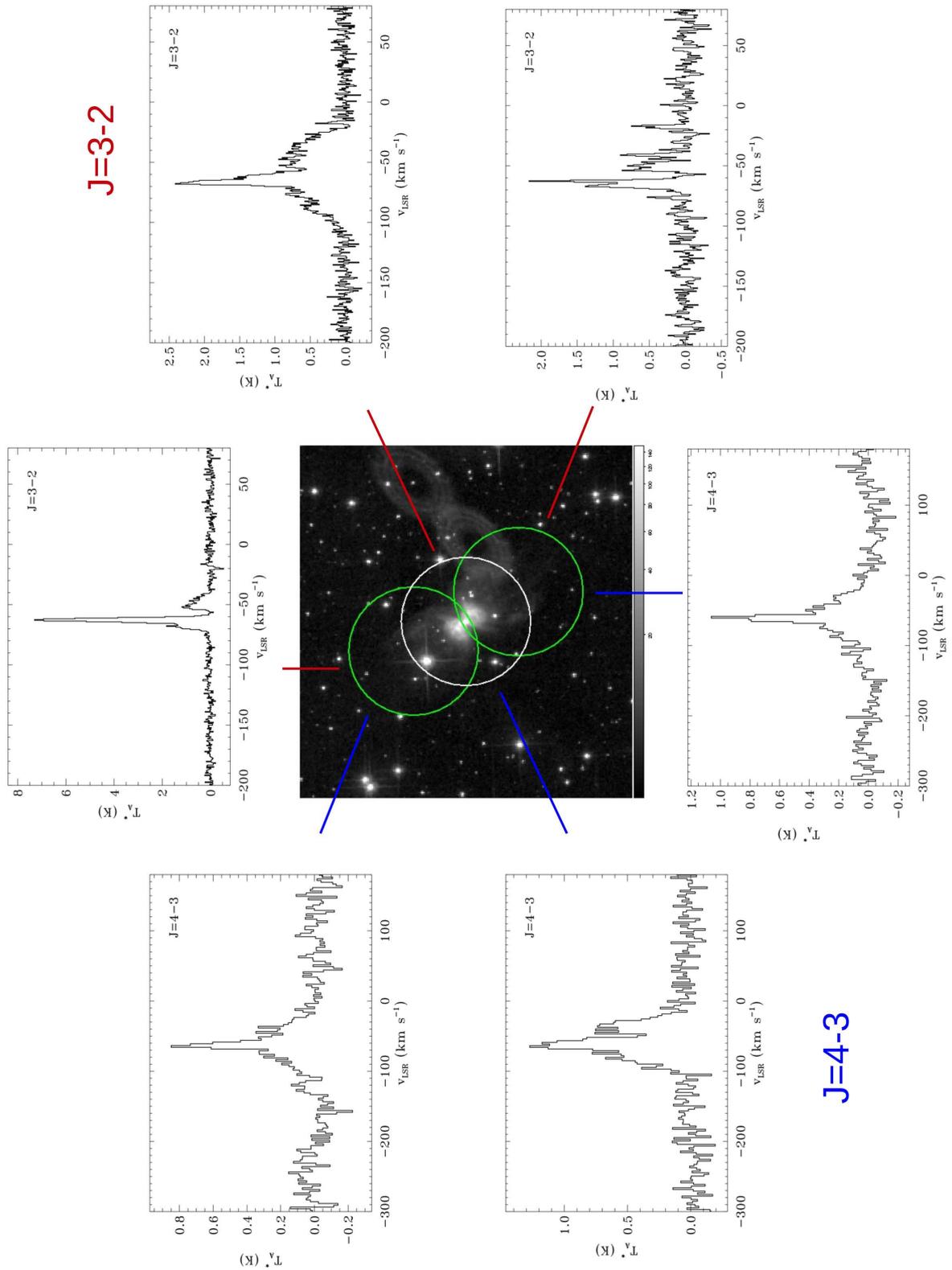}
\caption{Circles as wide as the APEX beam at 345 GHz are centered around our pointing coordinates: a white circle indicates the central pointing and two green circles indicate the two pointings $10''$ off-set. The I-band background image of I15452 was retrieved from the HST archive (North up and East left). Red bars indicate the spectra obtained for the $\mathrm{J}=3-2$ CO line and blue bars those for the $\mathrm{J}=4-3$ line.}
\label{fig:hst}
\end{figure*}

We have obtained three spectra corresponding to the central-star coordinates as derived from observations performed with the Hubble Space Telescope
 and two positions $7''$ off-set in both right ascension and declination. These are shown in Figure~\ref{fig:hst}.

\begin{figure}
 \centering
\includegraphics[width=0.4\textwidth]{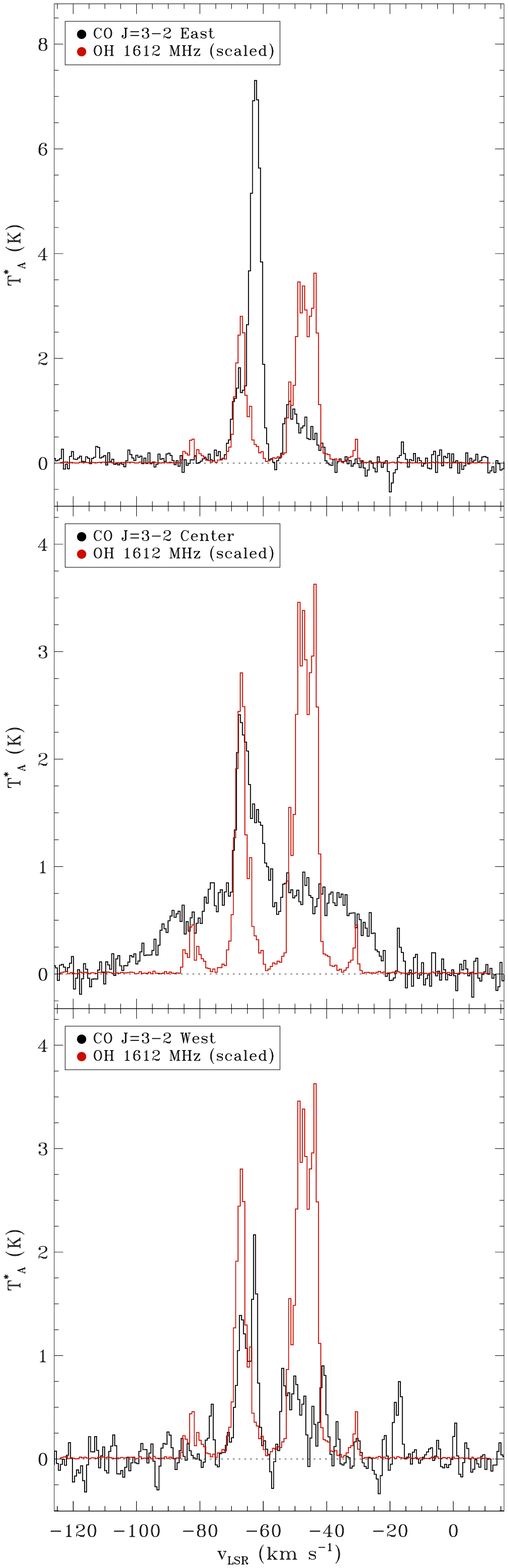}
\caption{Overlay of our CO $\mathrm{J}=3-2$ data and OH 1612 MHz \citep{deacon04} lines for IRAS~$15452-5459$. {The off-sets from the central position in (RA, DEC) for the Noth-East and South-West pointings were (7, 7) and ($-7$, $-7$), respectively. }}
\label{fig:overlay}
\end{figure}

The spectra corresponding to the central pointing appear as a superposition of at least two components: a broad feature and a
narrow one. The broad feature seems to show an asymmetric profile, its red side being relatively abruptly cut off.
Such a shape is likely due to the presence of an absorption feature around $-20$~km~s$^{-1}$ clearly detected 
in the off-set pointings, which is probably masking the line profile. The absorption was already noticed in the 
OH maser spectra by \citet{deacon04}, who also pointed out that a similar feature is present around $-55$~km~s$^{-1}$. These are likely
spurious features due to the ISM emission in the off-source position. In Figure~\ref{fig:overlay}, we compare the CO features with the OH maser spectrum published by \citet{deacon04} and notice that the wide CO component extends over a much larger velocity range than the maser features.
This broad component is not detected in the off-set pointings, which indicates that the part of
the outflow from which this feature originates is located in the inner region of the CSE. This is confirmed by the detection of the broad feature
in all pointings in the $\mathrm{J}=4-3$ line, which is a better tracer of the inner region.

In the $\mathrm{J}=3-2$ off-set pointings, we detect only a central, strongly asymmetric double-horn component, whose red feature 
is much weaker than the blue one. At a close inspection, the spectral structure looks more like a four-peak feature, resembling an optically
thin, resolved double shell.

\subsection{Kinematical components}
The data acquired North East and South West of the central star are characterized by a double-peak structure.
The blue peak shows a double-peak substructure that is easily seen also in the south-western pointing. An analogous substructure is probably responsible for the shape of the red peak. This closely resembles what {is}
 observed in the OH masers, where two features centered around $-57$~km~s$^{-1}$ also show multi-peak substructures.


By fitting Gaussians to the peaks, we derive for the NE feature a central velocity of $-57.5$~km~s$^{-1}$ and a velocity width (FWHM) 
of $27.4$~km~s$^{-1}$, while for the SW feature we obtain $-55.5$ and $28.6$~km~s$^{-1}$, respectively. 

As can be seen in Figure~\ref{fig:overlay}, the narrow peaks match very well with the velocity ranges of the two central features observed in the OH maser spectrum by \citet{deacon04}. They are also centered around the velocity accredited to the central star from the analysis of the OH maser lines. In spite of this, such narrow features cannot be conclusively distinguished from ISM contamination without carefully mapping the area around our target.

On the contrary, the broad component detected in the central pointing can be certainly attributed to I15452. We fit this component with a function of the form 
\begin{equation}
 T_{mb}(v)=T_{mb}^{peak}\left[1-\left(\frac{V-V_c}{V_e}\right)^2\right], \;\;|V-V_c|<V_e
\end{equation}
and null elsewhere. As central velocity $V_c$ we take $-57.5$~km~s$^{-1}$ and find $T_{mb}^{peak}\sim 0.8$~K and $V_e=45.0$~km~s$^{-1}$.


\begin{figure}
\centering
\includegraphics[width=0.4\textwidth,clip]{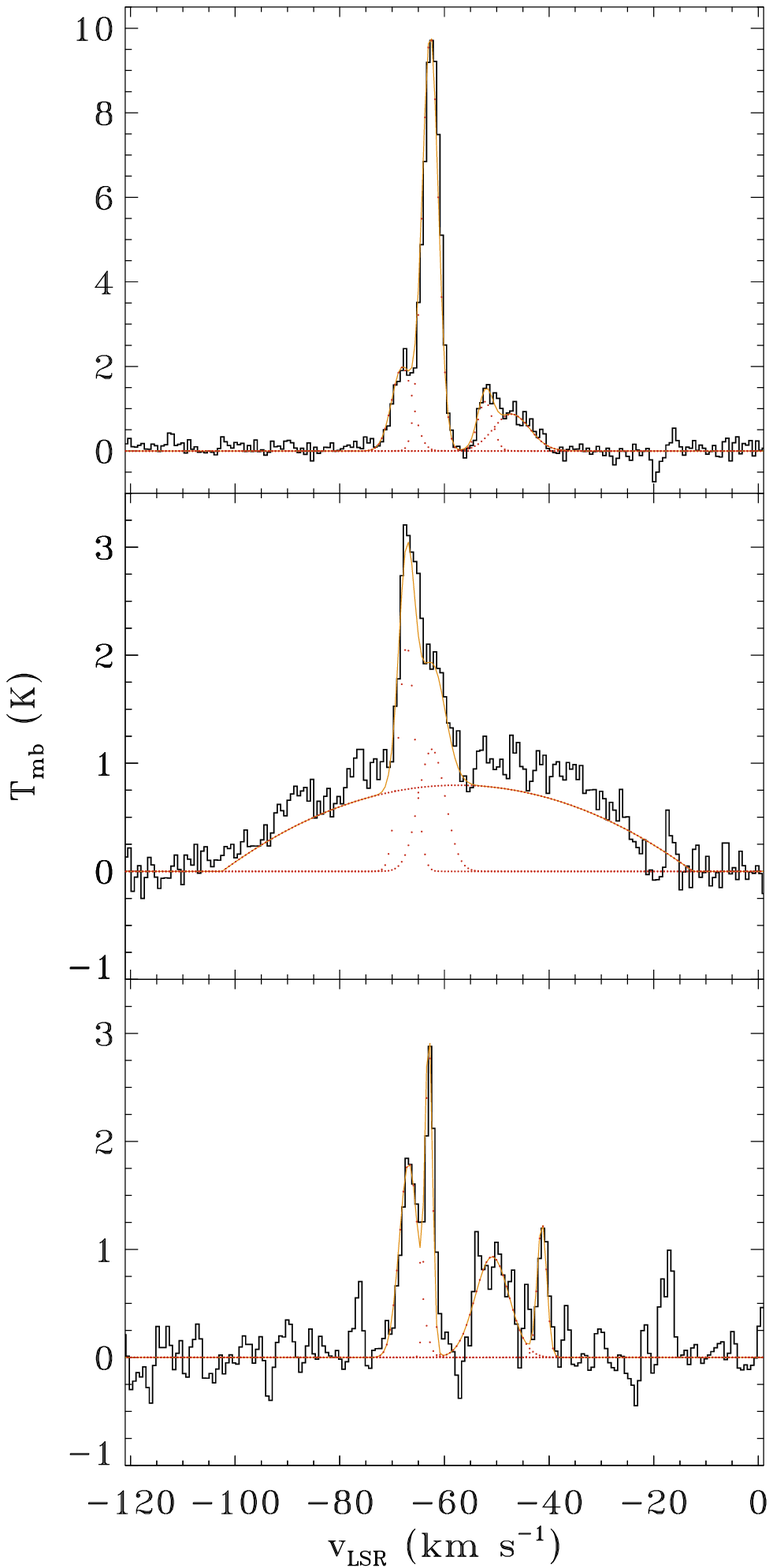}
\caption{Spectra of the $\mathrm{J}=3-2$ line observed $10''$ North-East of the central source in IRAS~$15452-5459$ (\textit{top}), on the central source
(\textit{center}) and $10''$ South-West of the central position (\textit{bottom}). Overplotted in red are Gaussian fittings
to the narrow features and a parabola for the broad component in the central pointing.}
\label{fig:gaussians}
\end{figure}

\begin{figure*}
 \centering
\includegraphics[width=0.9\linewidth,clip]{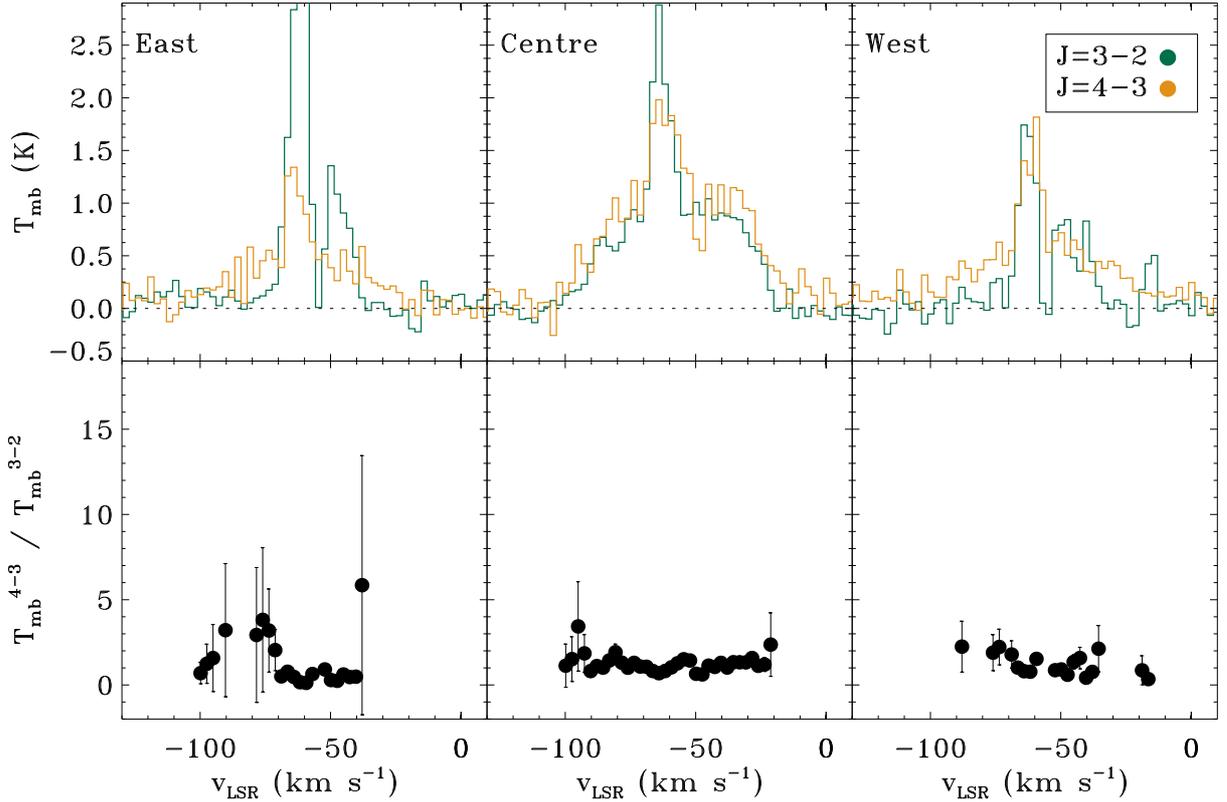}
\caption{\textit{Top:} The two CO lines detected are plotted on each other, after resampling the spectra at a resolution of 2.5~km~s$^{-1}$.
\textit{Bottom:} the ratio of the two lines is shown for points where the emission from both lines is above their respective 
noises.}
\label{fig:ratio}
\end{figure*}

\subsection{Mass-loss rate}
We do not have observations of $^{13}$CO lines that would allow us to estimate the optical depth of the $^{12}$CO lines.
We can calculate the mass-loss rate associated to the outflow generating the broad component from the correlation between this parameter and the $T_{mb}$ at the center of the line.

{For a spherically symmetric envelope, } if we assume that its emission is optically thin, we can estimate the mass-loss rate 
from \citet{knapp} and \citet{olofsson} as 
\begin{equation}
\dot{M}=4.55\times10^{-19}\left[\frac{T_{mb}}{\mathrm{Log}(W/0.04)s(J)}\right]^{5/6}\,f_{CO}^{-1}\,V_{exp}^{11/6}\,(D\,\theta)^{5/3} 
\label{eq:massloss}
\end{equation}
where $T_{mb}$ is the line peak main-beam temperature in K, W is the ratio of the flux emitted by the target at 4.6~$\mu$m and that
emitted by a star at the same distance with radius $5\times10^{13}$~cm approximated by a Planck curve at 2000 K, $s(J)$ is a correction factor
to account for the specific $\mathrm{J} \rightarrow \mathrm{J} -1$ transition, $f_{CO}$ is the abundance of CO relative to H$_2$, $V_{exp}$ the expansion
velocity of the outflow in km~s$^{-1}$, D the distance to the star in pc, and $\theta$ the beam size in arcsec.

From our fitting to the component, we derive $T_{mb}=0.8$~K and $v_{exp}=45$~km~s$^{-1}$. To
 estimate the W parameter, we perform synthetic photometry from the observed ISO spectrum of the source
and calculate the IRAC 4.5~$\mu$m flux density with the Spitzer data reduction package SMART\footnote{SMART was 
developed by the IRS Team at Cornell University and is available through 
the Spitzer Science Center at Caltech.} \citep{smart}, obtaining 11.38 Jy (the source
is included in the GLIMPSE survey \citep{glimpse}, but its photometry at 4.5~$\mu$m is not available through the on-line GLIMPSE catalog, probably because of saturation).
We thus obtain $\mathrm{W}=0.75$. The correction factor $s(3)$ is taken as 0.43, following \citet{groenewegen} and $f_{CO}=3\times10^{-4}$, which is typical
 of O-rich circumstellar envelopes. Finally, the distance is set to 2.5 kpc and the beam is 17.3$''$. The derived mass loss rate 
is $1.2\times10^{-4}$~M$_\odot$~yr$^{-1}$. 

Under the assumption of an optically thick line, the mass loss rate can be estimated as \citep{groenewegen}
\begin{equation}
 \dot{M}=1.4\frac{T_{mb}\,V_e^2\,D^2\,\theta^2}{2\times10^{19}\, f_{CO}^{0.85}\, s(J)}
\end{equation}
which returns a value of $4.9\times10^{-4}$~M$_\odot$~yr$^{-1}$. In both cases we obtain quite high values, but the calculations assume spherical symmetry,
therefore their results are probably overestimated.

Still distinguishing between optically thin and thick regimes, we can calculate the mass of molecular gas corresponding to each spectral component.
The H$_2$ masses were derived by integrating over the whole spectrum the following formula for the $i$-th channel
\begin{equation}
 M^{H_2}_i=m_{H_2} D^2 \Omega N^{CO}_i f_{CO}
\end{equation}
where the CO column density $N_i^{CO}$ is given by  \citet{choi} as
\begin{equation}
N^{CO}_i=1.1\times10^{15}\frac{T_{mb}\Delta v}{D(n,T_K)}\frac{\tau_{32}}{1-\textnormal{exp}(-\tau_{32})}
\end{equation}

$D(n,T_K)$ was shown to be equal to 1.5 within a factor of 2 in the range $10<T_K<200$~K and $10^4<n<10^6$~cm$^{-3}$, $m_{H_2}$ is the mass
of a molecule of hydrogen, $D$ the distance to the star in cm, and $\Omega$ the beam in sr.
In Table~\ref{tab:parameters32}, we list values for $\tau_{32}\ll 1$ and for $\tau=5$ as a fiducial for the optically thick case, when indeed the mass depends on the real value of $\tau$.
We also calculated the energy and momentum associated to the outflow along the line of sight by 
integrating $P_i=M_i|v_i-v_0|$ over the spectrum (the central velocity $v_0$ is $-57.5$~km~s$^{-1}$) and then $E=\frac{P^2}{2M}$ . 

The values of momentum and energy can be compared with those found by \citet{bujarrabal}. They find that most of the standard proto-PNe show values of kinetic energy of their outflows in the range $10^{44}$--$10^{46}$~erg, with low-mass stars and hypergiants respectively below and above this range. 
Depending on the opa\-ci\-ty of the features observed, we find in the molecular envelope around I15452 kinetic energies in the range $10^{42}$--$10^{44}$~erg, which would then imply its classification as a low-mass star.

\citet{bujarrabal} also find that if the origin of the outflow is radiation pressure on grains, an upper limit to the scalar momentum is given by $P\lesssim L/c$~(g~cm~s$^{-1}$~yr$^{-1}$)$~\times 2000$~(yr). For our source, $L/c=3.24\times10^{34}$~g~cm~s$^{-1}$~yr$^{-1}$, therefore the upper limit to $P$ is $6.48\times10^{36}$~g~cm~s$^{-1}$. We obtain for the momentum $6.96\times10^{36}$~g~cm~s$^{-1}$ under the  optically-thin assumption. Even in this conservative case, the momentum is larger than what can be afforded by the radiation: only action of radiation on grains cannot explain the outflow  we observe in our target, as has already been found by \citet{bujarrabal} for many proto-PNe.

\begin{table}[ht]
\centering
 \begin{tabular}{lccc}
\hline
\hline\noalign{\smallskip}
    $\dot{\textnormal{M}}$                   & M$_{H_2}$  &       $P$         &     $E$     \\

                 10$^{-4}$ M$_\odot$~yr$^{-1}$ & $10^{-3}$M$_\odot$          & $10^{36}$~g~cm~s$^{-1}$ & 10$^{42}$~erg  \\
\hline\noalign{\smallskip}
\multicolumn{4}{c}{$\tau\ll 1$} \\
 1.2                           & 2.1               & 6.96                  & 5.80 \\
 \multicolumn{4}{c}{$\tau=5$} \\
 4.9                           & 80               & 284                 & 254       \\
\hline
 \end{tabular}
\caption{Physical parameters derived for the outflow from the $\mathrm{J}=3-2$ line.}
\label{tab:parameters32}
\end{table}


\section{IRAS~$15452-5459$: IR archive observations}
The star has been imaged with the Hubble Space Telescope (HST) at optical (V) and near-IR (I, J, H, and K) wavelengths by \citet{sahai}. The V-band images do not show any ne\-bu\-lo\-si\-ty and only the central star is weakly detected. The nebula becomes brighter at longer wavelengths, exhibiting an hourglass shape ($\sim25''$), 
 as described by \citet{sahai}, to which we refer for a morphological description. 


\subsection{The ISO spectrum}
I15452 was observed with the Short and Long Wavelength Spectrometers on-board the Infrared Space Observatory (ISO). Its IR spectrum was shown by \citet{garcialario},
who reported the detection of the mid-IR features attributed to Polycyclic Aromatic
 Hydrocarbons (PAHs) at 6.2, 7.7, and 8.6 $\mu$m, a plateau between 10 and 15 $\mu$m also attributed to PAHs, and the $30 \mu$m
feature due to MgS. All of these features are typical of C-rich environments. 
The presence of both PAH features and the masers
would make this object a member of the rare class of \textit{mixed-chemistry} sources, whose nature is not well understood. These may be
stars that have recently turned their chemistry from O-rich into C-rich and during the transition show properties of both
CSEs, or they may host structures that retain O-rich material, in spite of the chemical change in the overall CSE. Another option
is that photodissociation makes atomic C and O available for reaction and then both O-bearing and C-bearing chemical species are
found in the CSE \citep{little}.

\begin{figure}
\centering
  \includegraphics[width=0.45\textwidth,clip]{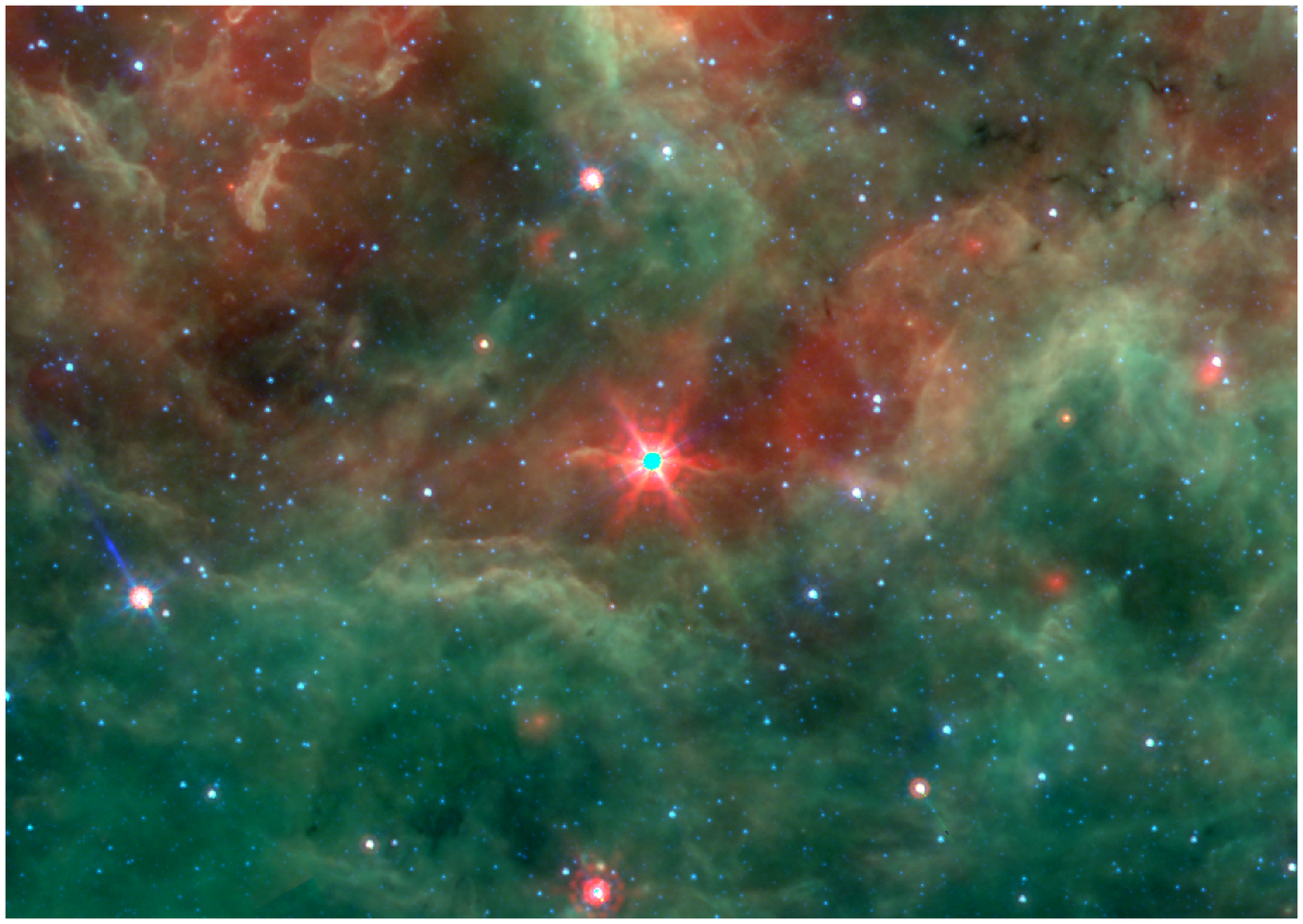}
\caption{Spitzer composite of the region around I15452 from the GLIMPSE and MIPSGAL archives. Images at 5.8, 8.0 and 24 $\mu$m are coded on a logarithmic scale as blue, green and red, respectively. The image covers an area about $24^\prime \times 19'$ wide. North is up and East left.}
\label{fig:spitzer}
\end{figure}

The spectrum also shows several atomic forbidden emission lines of [OIII], [NIII], [NII],
 [CII], and [OI], which are typical of photodissociation regions (PDRs), but the continuum is clearly contaminated by spurious emission that can be attributed to Galactic cirrus.
We have retrieved the highly-processed data set from the ISO  web archive and in Table~\ref{tab:lines} we list the  lines
 detected in the spectrum, their widths and their fluxes.
\begin{table*}[ht]
\centering
\begin{tabular}{lccc}
\hline\hline\noalign{\smallskip}
Line & Peak Wavelength & FWHM & Flux \\
   & $\mu$m        & $\mu$m & $10^{-18}$ W cm$^{-2}$ \\ 
\hline\noalign{\smallskip}
$[$CII$]$ (157.74)	&$157.724 \pm 0.002	$&$0.662 \pm 0.002$	&$5.42 \pm 0.04 $\\
$[$OI$]$ (145.52)	&$145.54 \pm 0.03	$&$0.65 \pm 0.03$	&$0.16 \pm 0.02 $\\
$[$NII$]$ (121.90)	&$121.97 \pm 0.02	$&$0.99 \pm 0.02	$&$2.0 \pm 0.1 $\\
$[$OIII$]$ (88.36)	&$88.367 \pm 0.008	$&$0.316 \pm 0.007$	&$3.6 \pm 0.2  $\\
$[$OI$]$ (63.18)	&$63.191 \pm 0.007	$&$0.24 \pm 0.07$	&$0.79 \pm 0.07  $\\
$[$OIII$]$ (51.81)	&$51.79\pm0.02		$&$0.45 \pm 0.02	$&$3.9 \pm 0.6  $\\ 
$[$NIII$]$ (57.33)	&$57.32 \pm 0.01	$&$0.26 \pm 0.01	$&$1.7 \pm 0.2  $\\
$[$SiII$]$ (34.84)	&$34.793 \pm 0.003	$&$0.09 \pm0.007 $	&$0.43 \pm 0.03$ \\
$[$SIII$]$ (33.48)	&$33.46 \pm 0.01	$&$0.160 \pm 0.009$	&$2.4 \pm 0.4$ \\
\hline  \end{tabular}
\caption{Atomic lines detected in the ISO spectrum of I15452. In co\-lumn one, the expected peak wavelengths are listed in $\mu$m.}
\label{tab:lines}
\end{table*}
 Indeed, as can be seen in Figure~\ref{fig:spitzer}, the area where the source is located is rich in molecular gas and dust, and, since a background spectrum is not available, it cannot be ruled out that this diffuse gas affects or is responsible for the spectral features observed, keeping in mind that the FWHM of the ISO beam was  $\sim 1.5''$--$93''$ in the 2--245~$\mu$m range.

\subsection{PDR lines and circumstellar radiation field}
In a study of far-IR  PDR lines in AGB stars, proto-PNe, and PNe, \citet{fong} and \citet{castrocarrizo} notice that the detection of PDR lines
depends mostly on the temperature of the central star, and that shocks, massive outflows, and the interstellar radiation
 field do not seem to play a role.
 On the basis of their observations, they distinguish circumstellar from interstellar PDRs by the flux ratio of the $[$CII$]$ (158 $\mu$m) to $[$OI$]$ (63 $\mu$m) lines. The authors in fact observe that the former is always less intense than the latter, when both {lines} are clearly attributed to a circumstellar PDR. This clear attribution was not possible for their coldest sources, for which background spectra {indicated interstellar contamination.} This can bias the conclusion that in circumstellar PDRs I$_{[\mathrm{CII}]}$ can be expected to be smaller than I$_{[\mathrm{OI}]}$.
 
\begin{figure}
 \centering
\includegraphics[width=0.45\textwidth]{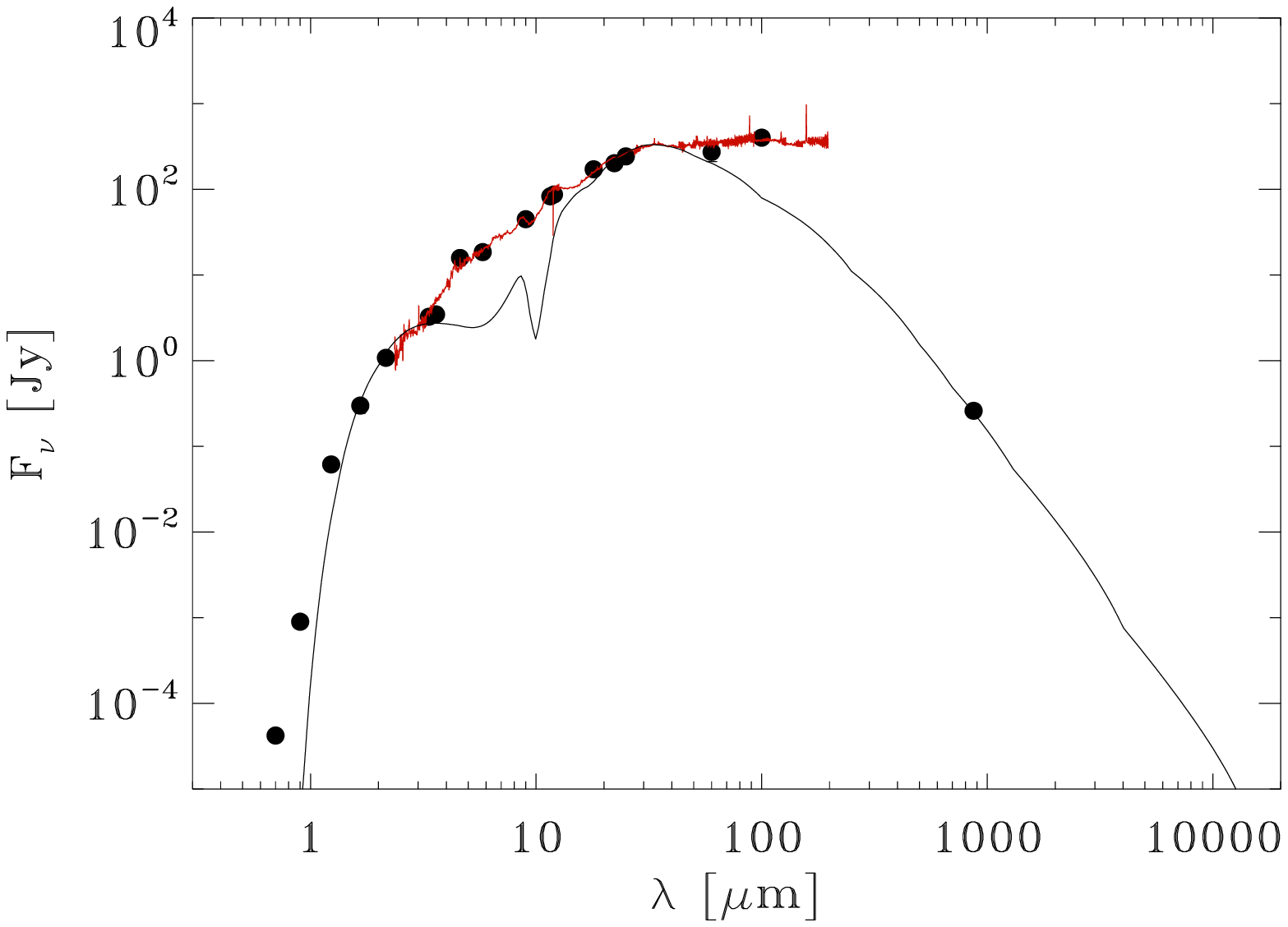}
\caption{SED of I15452 with the ISO spectrum shown in red and the DUSTY model as a black solid line. {The data from Table~\ref{tab:catalogs} and our measurement at 345 GHz (870~$\mu$m) are shown as solid circles.}}
\label{fig:sed}
\end{figure}

 {
To characterize the average radiation field around the star, we can introduce the parameter $G$, the intensity of the UV radiation field in Habing units \citep{castrocarrizo}:
 \begin{equation}
G=\frac{LF_{UV}}{4\pi R_i^2 G_0} 
\end{equation}
where $L$ is the stellar luminosity, $F_{UV}$ the fraction of energy at wavelengths shorter than 6~eV ($\sim 2066$~\AA) in a blackbody spectrum at the same temperature as the star, $R_i$ is the inner radius from the star to the PDR and $G_0=1.6\times10^{-3}$~erg~cm$^{-2}$~s$^{-1}$ is the average UV flux in the diffuse ISM.

  The ratio between $[$CII$]$ (158 $\mu$m) and $[$OI$]$ (63 $\mu$m) lines depends on the physical conditions in the PDR and the $[$CII$]$ line is known to be the dominating coolant for average radiation fields with $G\lesssim10^3$, while the $[$OI$]$ dominates at higher values of G. Therefore, although contamination is certainly possible, the atomic lines observed in the ISO spectrum of I15452 may indeed arise in a circumstellar PDR with a weak average radiation field.

We can calculate what value of G we expect for a star like I15452.
 The estimation of $R_i$ can be done by SED modeling, taking as $R_i$ the inner radius of the dust shell.
To model the SED,
}
 we have collected photometric data from several catalogs and used the code DUSTY \citep{dusty} to solve the equation of radiation transfer in the circumstellar dust shell. DUSTY only treats spherical symmetry, which is clearly not our case, but it can still provide us with an approximate estimate of the inner radius.  
In Table~\ref{tab:catalogs}, we have listed the data collected from on-line catalogs, namely USNOB1, 2MASS, GLIMPSE, WISE, AKARI and IRAS; the data are shown in Figure~\ref{fig:sed}.

\begin{table*}\centering
 \begin{tabular}{llcccc}
\hline\hline\noalign{\smallskip}
 \multirow{2}{*}{USNOB1} & $\lambda$ $[\mu$m$]$ & {0.7} & {0.9} & & \\
                         & F$_\lambda$ $[$mJy$]$ & {$0.042\pm0.004$} & {$0.90\pm0.09$} & & \\
\hline\noalign{\smallskip}
 \multirow{2}{*}{2MASS} & $\lambda$ $[\mu$m$]$ &  1.235 & 1.662 & 2.159 & \\
                           & F$_\lambda$ $[$mJy$]$ &  $61.5\pm8.7$ & $299\pm56$ & $1080\pm30$ & \\
\hline\noalign{\smallskip}
 \multirow{2}{*}{IRAC} & $\lambda$ $[\mu$m$]$ & {3.6} & {5.8} & & \\
                       & F$_\lambda$ $[$Jy$]$ &{$3.470\pm0.49$} & {$18.52\pm0.76$} & & \\
\hline\noalign{\smallskip}
 \multirow{2}{*}{WISE} & $\lambda$ $[\mu$m$]$ & 3.35 & 4.6 & 11.56 & 22.24 \\
                       & F$_\lambda$ $[$Jy$]$ & $3.26\pm0.37$ & $15.81\pm0.73$ & $82.77\pm0.61$ & $201.8\pm3.7$ \\
\hline\noalign{\smallskip}
  \multirow{2}{*}{AKARI} & $\lambda$ $[\mu$m$]$ & {9.0} & {18.0} & & \\
                            & F$_\lambda$ $[$Jy$]$ & {$44.8\pm3.9$} & {$171.4\pm10.7$} & & \\
\hline\noalign{\smallskip}
  \multirow{2}{*}{IRAS} & $\lambda$ $[\mu$m$]$ & 12 & 25 & 60 & 100\\
                        & F$_\lambda$ $[$Jy$]$ & $87\pm9$ & $243\pm12$ & $274\pm63$ & 401 \\
\hline
\end{tabular}
\caption{Photometric data on I15452 collected from on-line catalogs. {Infrared data are available on line through the IRSA archive (http://irsa.ipac.caltech.edu), while optical data are available through VizieR \citep[http://vizier.u-trasbg.fr/viz-bin/VizieR]{vizier}}.}
\label{tab:catalogs}
\end{table*}

We ran our model with T$_\star=3500$~K, standard density distribution of r$^{-2}$, grain size distribution with minimum radius $a_{min}=0.005$ and maximum $a_{max}=0.25$~$\mu$m varying as $a^{-0.35}$ and optical constants for amorphous silicates (appropriate for O-rich environments). With the exception of the mid-IR range, where the excess of emission can be regarded as an indication for a disk, a good fit to the observational data can be obtained with a distance of 2.5 kpc, a luminosity of 8000~L$_\odot$, and {a circumstellar absorption} $A_V=58$ ($\tau\sim1$ around 3.35~$\mu$m). 

From our DUSTY modeling, we can estimate a FIR continuum flux of $1.7\times10^{-11}$ W~m$^{-2}$ and an inner radius of $5.7\times10^{15}$~cm. 
Therefore we find  $\mathrm{G} \sim 170$, with $F_{UV}=3.4\times10^{-6}$. I15452 seems then to fall within the regime where the $[$CII$]$ line is the dominant coolant.

To estimate the intensity of the FUV field and the density in the PDR, we use the models described in \citet{kaufman} and available on-line at http://dustem.astro.umd.edu/. For a given set of gas-phase elemental abundances and grain properties, each model is described by a constant H nucleus density, $n$, and incident far-ultraviolet intensity $G$. The models solve for the equilibrium chemistry, thermal balance, and radiation transfer through a PDR layer.  The values of $n_H$, and $G$ are obtained by overlaying two sets of contour plots and finding the intersection of the observations. The procedure requires either two sets of line-intensity ratios or one set of line-intensity ratios and the ratio of the $[$CII$]+[$OI$]$ intensity to that of the infrared dust continuum emission. {The abundances used in the calculation  were $[$C/H$] = 1.4 \times 10^{-4}$, $[$O/H$] = 3 \times 10^{-4}$, $[$Si/H$] = 1.7 \times 10^{-6}$, $[$S/H$] = 2.8 \times 10^{-5}$, $[$Fe/H$] = 1.7 \times 10^{-7}$, $[$Mg/H$] = 1.1 \times 10^{-6}$ and $[$PAH/H$] = 4 \times 10^{-7}$.}

Either a filling-factor correction should be applied to each line  or line intensities should be used, as lines of different species do not necessarily arise from the same regions. For example, diffuse $[$CII$]$ emission may fill the observing beam, whereas $[$OI$]$ emission may arise from a smaller region at higher density and temperature. A similar correction should also be applied to the FIR continuum.

\begin{figure}
 \centering
\includegraphics[width=0.45\textwidth]{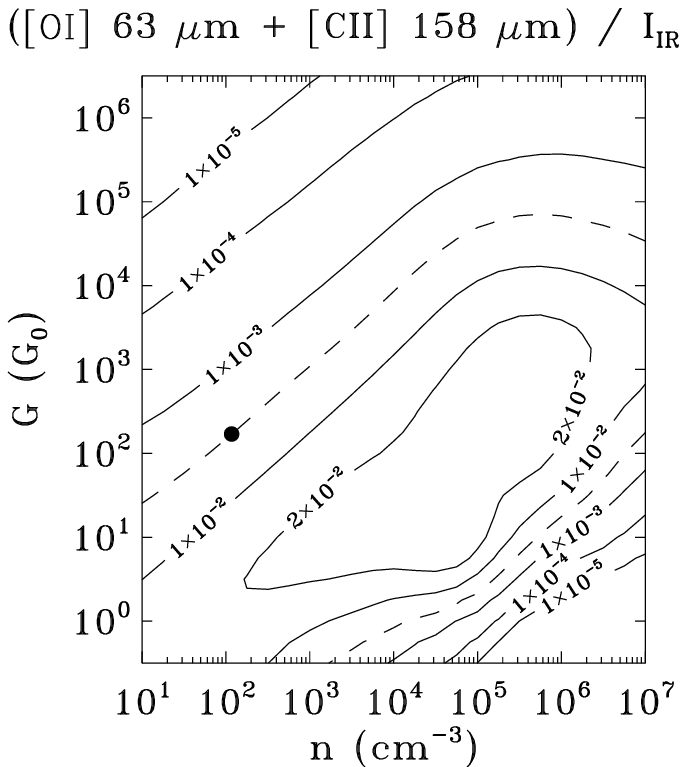}
\caption{Contour plots of  {the ratio of the intensity of the $[$CII$]$ line plus that of the $[$OI$]$ line over the FIR continuum as a function of the intensity of the FUV field $G$ and the atomic H density $n$. The dashed contour corresponds to the value of 0.0037 found for I15452 and the solid circle corresponds to $G = 170$ and $n = 117$.}}
\label{fig:ratios}
\end{figure}

In Figure~\ref{fig:ratios}, we show contour plots for the ratios between the sum of the two PDR lines and the FIR continuum as a function of $G$ and $n$, obtained from the PDR modeling.
In the rough assumption that all of the emitting components arise from regions with similar sizes, we have $(I_{[OI]}+I_{[CII]})/I_{FIR}\sim0.0037$. For our calculated value of $G$, this corresponds in Figure~\ref{fig:ratios} to a density of 117~cm$^{-3}$. This density value is somewhat low for a post-AGB CSE, but may be found in the tenuous region where the fast wind has swept away most of the material.

\subsection{Envelope mass}
 We can derive the mass of dust by simply fitting a modified blackbody to the observational data beyond 20~$\mu$m, neglecting the IRAS point at 100~$\mu$m, because clearly contaminated.
This procedure provides us with an average dust temperature T$_\mathrm{d}$ and a dust emissivity index $\beta$ that can be used to calculate  the mass of dust following \citet{beuther}:
\begin{equation}
\lefteqn{M_{dust}=\frac{2.0\times 10^{-4}}{J_\nu(T_{d})}\frac{F_\nu}{\mbox{Jy}}\left(\frac{D}{\mbox{kpc}}\right)^2\left(\frac{\nu}{1.2 \; \mbox{THz}}\right)^{-3-\beta} \; M_\odot}   
\label{eq:dustmass}
\end{equation}
\[
\lefteqn{J_\nu(T_d)=\left[\mbox{exp}\left(\frac{h\nu}{kT_d}\right)-1\right]^{-1} } 
\]

The formula assumes an average grain size of 0.1~$\mu$m and grain density of 3~g~cm$^{-3}$;   $h$ and $k$ are the Planck and Boltzmann constants, D is the distance to the star (2.5 kpc), $F_\nu$ the flux density measured at the frequency $\nu$, which in our case is that of our LABOCA measurement, 345 GHz. 

The fit to the far-IR/sub-mm observational points gives us T$_\mathrm{d} \sim 95 \pm 9$~K and $\beta \sim 1.10 \pm 0.14$, which allows us to calculate a CSE dust mass of $0.010 \pm 0.003$~M$_\odot$. The uncertainties listed are the statistical errors resulting from the fitting procedure and from error propagation in Eq.~\ref{eq:dustmass}.

\citet{castrocarrizo} derive a formula to estimate the mass of atomic gas from the flux
observed in the $[$CII$]$ line at 157.8~$\mu$m, which we report here:
\begin{equation}
 M(M_\odot)=7.0\times10^6\, F_{[CII]}(\mathrm{erg~cm}^{-2} s^{-1})\, D(\mathrm{kpc})^2\, F_c/X_C
\label{eq:atomicmass}
\end{equation}
where $F_{[CII]}$ is the observed flux in the $[$CII$]$ line, $D$ the distance to the star, $X_C$ the abundance of C ($1.5\times$10$^{-3}$ for C-rich stars and 
$2\times$10$^{-4}$
for O-rich sources), and $F_c$ is a correction factor that accounts for low excitation temperatures. The correction factor is
always $\geq1$ and depends on the density and kinetic temperature of the gas, but at the densities found in proto-PNe it is
unlikely to be larger than 2 or 3.

From Eq.~\ref{eq:atomicmass}, we can then calculate the mass of atomic gas in the nebula around  I15452
to be about 1.9~M$_\odot \;D$(kpc)$^{2}$,  assuming $F_c=1$. I15452 appears to be surrounded by a massive CSE: still assuming $F_c=1$, it would contain about 12~M$_\odot$ of gas, at the distance of 2.5 kpc. {A statistical error on the mass estimate may be calculated from the error on the flux of the $[$CII$]$ line, but this would not account for the real uncertainty on the mass, which is much larger and due to the degree of contamination.} The high value of mass found is beyond what is expected from intermediate-mass stars and contrasts with the small momentum found above for the molecular component, which is considered as an indication for the source to be a low-mass star. Such a large estimate for the circumstellar atomic mass can be regarded as an indication for ISM contamination of the far-IR spectrum. 

{For a gas-to-dust mass ratio of 200 - a standard value in evolved stars - our dust mass estimate implies a gas mass of about 2~M$_\odot$. Contamination would then enhance the flux in the $[$CII$]$ line by a factor of 6. }

\section{Discussion and conclusions}
We have observed a sample of stars where the detection of maser velocities larger than what observed on the AGB indicates that these sources have started their post-AGB evolution. In particular, we have included four of the water-fountain nebulae known that were not previously investigated for CO emission. As our targets are located in the Galactic plane, spurious emission from the interstellar medium typically contaminates the spectra and does not allow us to obtain conclusive detections. By comparing the maser features with the CO lines, we find hints for possible CO detections, but only with sensitive maps of the regions around the targets can a distinction between ISM and circumstellar CO be made.

Only IRAS~$15452-5459$ is clearly detected in both the $\mathrm{J}=3-2$ and $\mathrm{J}=4-3$ lines. The star shows in its central region a fast CO outflow with velocities of about $V_{exp}=45$~\kms, while narrow peaks  point to an expansion velocity of $\sim14$~\kms. The latter falls within the range of velocities observed during the AGB and, as it is detected at all pointings, it can be interpreted as due to the remnant AGB wind. With its four peaks, the CO feature resembles what observed in  the OH maser and points to a complex emitting structure, rather than an expanding spherical envelope. 
The broad component observed only in the central region of the source points instead to an outflow of molecular gas. If such an outflow is in the form of a collimated wind can be estimated only with interferometric observations. The mass-loss rate calculated for this component is quite high ($\sim 10^{-4}$~M$_\odot$~yr$^{-1}$) but the assumed spherical symmetry implies that the value is likely overestimated. We have also calculated the scalar momentum and kinetic energy associated to the outflowing material, which we have compared with the results obtained by \citet{bujarrabal}, concluding that radiation pressure on dust grains alone cannot explain the dynamics of the CSE.

{ By fitting a modified blackbody to the far-IR/sub-mm observational points of I15452, we have estimated a dust mass in its CSE of $\sim$0.01~M$_\odot$, while through the analysis of its ISO spectrum, we have calculated an envelope gas mass of about 12~M$_\odot$. The uncertainty in the latter value is dominated by the degree of contamination of the $[$CII$]$ line. 
 The derived values of gas and dust masses (12 and 2 M$_\odot$, respectively) indicate an enhancement factor of 6 of the line flux because of con\-ta\-mi\-nation, if we assume a standard value of 200 for the gas-to-dust mass ratio.

Nevertheless, the total CSE mass of I15452 may be larger than 2~M$_\odot$ and a higher gas-to-dust mass ratio should be consi\-dered.}
I15452 belongs to a group of OH masing stars classified by \citet{sevenster} as LI, because they lie on the left of the evolutionary sequence on the IRAS color-color diagram, while \lq\lq standard\rq\rq~post-AGB stars lie on the right side (RI sources).   LI stars have higher outflow velocities and mass-loss rates and they may loop back to the left of the evolutionary sequence because of an early AGB termination, which would also cause them to host OH masers for a much longer time than RI stars \citep{sevenster}. Finally, by scale height considerations, \citet{sevenster} argue that LI stars may be the precursors of bipolar planetary nebulae and have larger masses than RI sources (ending in elliptical planetary nebulae), with average values of 4 and 1.7~M$_\odot$ for LI and RI, respectively \citep{deacon07}. Indeed, the bipolarity of the nebula around I15452 has been proven by the Hubble images. {For this star to have a CSE mass within va\-lues adequate for LI sources (a minimum of 3--4~M$_\odot$), its gas-to-dust mass ratio must be  at least 300--400, on the basis of our dust mass estimate. This would decrease the degree of contamination of the $[$CII$]$ line mentioned above to a factor of 3--4.}

I15452 is certainly a peculiar object under many points of view. For example, the 4 peaks seen in its satellite OH line are almost unique among the several hundreds of evolved stars where this feature has been detected. Another target with such a profile of its OH maser is OH~$19.2-1.0$, in which the spectral shape of the features shows a smaller degree of symmetry than that in I15452. The four peaks in I15452 are stunningly similar to one of the profiles derived theoretically by \citet{grinin} in the case of an unsaturated maser originating from a rotating and expanding disk. As pointed out by \citet{chapman} for OH~$19.2-0.1$, the applicability of the calculations by \citet{grinin} to an evolved star is dubious, because OH masers in these sources are typically saturated. Nevertheless, the mid-IR excess of its SED and the mixed chemistry may actually point to the existence of a disk in I15452. The detection of the PAH features in a region of the ISO spectrum that does not appear to be affected by ISM contamination and the presence of O-bearing molecules can in fact be explained by a circumstellar disk. Such disk would serve as a reservoir of O-rich material, where the masers would arise, while the PAHs would be located in the circumstellar envelope, converted into C-rich by the third dredge-up.

Observations at high angular resolution of the mid-IR spectrum, the masers and the CO outflow are necessary to reconstruct the morphology of the central emitting region in I15452.

\begin{acknowledgements}
We would like to thank our referee, Dr Jessica Chapman, for comments that helped to improve this paper. This research has made use of the VizieR catalog access tool, CDS, Strasbourg, France. This research has made use of the NASA/ IPAC Infrared Science Archive, which is operated by the Jet Propulsion Laboratory, California Institute of Technology, under contract with the National Aeronautics and Space Administration. This research has made use of the VizieR catalogue access tool, CDS, Strasbourg, France.
\end{acknowledgements}


\begin{thebibliography}{}

\bibitem[Beuther et al, 2005]{beuther}
Beuther, H., Schilke, P., Menten, K. M., Motte, F., Sridharan, T. K., \& Wyrowski, F., 2005, ApJ, 633, 535

\bibitem[Boboltz \& Marvel(2005)]{bobo} 
Boboltz, D.~A., \& Marvel, K.~B. 2005, \apjl, 627, L45 

\bibitem[Bujarrabal et al.(2001)]{bujarrabal}
Bujarrabal, V., Castro-Carrizo, A., Alcolea, J., S{\'a}nchez-Contreras, C. 2001, A\&A, 377, 868

\bibitem[Carey et al.(2009)]{mipsgal}
Carey, S. et al. 2009, PASP, 121, 76

\bibitem[Castro-Carrizo et al.(2001)]{castrocarrizo}
Castro-Carrizo, A., Bujarrabal, V., Fong, D., Meixner, M., Tielens, A. G. G. M.,  Latter, W. B., Barlow, M. J. 2001, A\&A, 367, 674

\bibitem[Cerrigone et al.(2009)]{cerrigone}
Cerrigone, L., Hora, J. L., Umana, G., Trigilio, C. 2009, ApJ, 703, 585

\bibitem[Chapman, 1988]{chapman}
Chapman, J., 1988, MNRAS, 230, 415

\bibitem[Choi et al.(1993)]{choi}
Choi, M., Evans, N. J., Jaffe, D. T. 1993, ApJ, 417, 624

\bibitem[Churchwell et al.(2009)]{glimpse}
Churchwell, Ed et al. 2009, PASP, 121, 213

\bibitem[Day et al.(2010)]{day}
Day, F. M., Pihlstr{\"o}, Y. M., Claussen, M. J., Sahai, R. 2010, ApJ, 713, 986

\bibitem[Deacon et al.(2004)]{deacon04}
Deacon R. M., Chapman, J. M., Green A. J. 2004, ApJS, 155, 595

\bibitem[Deacon et al.(2007)]{deacon07}
Deacon R. M., Chapman, J. M., Green A. J., Sevenster, M. N. 2007, ApJ, 658, 1096

\bibitem[Engels(2002)]{engels}
Engels, D. 2002, A\&A, 388, 252

\bibitem[Fong et al.(2001)]{fong}
Fong, D., Meixner, M., Castro-Carrizo, A., Bujarrabal, V., Latter, W. B., Tielens, A. G. G. M., Kelly, D. M., Sutton, E. C. 2001, A\&A, 367, 652

\bibitem[Garc{\'{\i}}a-Lario et al.(2000)]{garcialario}
Garc{\'{\i}}a-Lario, P., Manchado Torres, A., Ulla, A., Manteiga, M., Su{\'a}rez Fernandez, O., 2000 ASPC, 199, 391

\bibitem[G{\'o}mez et al.(1994)]{gomez} G{\'o}mez, Y., 
Rodr{\'{\i}}guez, L.~F., Contreras, M.~E., \& Moran, J.~M.\ 1994, \rmxaa, 28, 97 

\bibitem[Grinin \& Grigor'ev(1983)]{grinin}
Grinin, V. P. \& Grigor'ev, S. A., 1983, Astron. Zh., 60, 512

\bibitem[Groenewegen et al.(1999)]{groenewegen}
Groenewegen, M. A. T., Baas, F., Blommaert, J. A. D. L., Stehle, R., Josselin, E., Tilanus, R. P. J. 1999, A\&AS, 140, 197 

\bibitem[G{\"u}sten et al.(2006)]{apex}
G{\"u}sten, R., Nyman, L. \AA., Schilke, P., Menten, K., Cesarsky, C., Booth, R. 2006 A\&A, 454, L13

\bibitem[Habing(1996)]{habing}
Habing, H. J. 1996, A\&AR, 7, 97

\bibitem[He et al.(2008)]{he}
He, J. H., Imai, H., Hasegawa, T. I., Campbell, S. W., Nakashima, J. 2008, A\&A, 488L, 21 

\bibitem[Heyminck et al.(2006)]{flash}
Heyminck, S., Kasemann, C., Güsten, R., de Lange, G., Graf, U. U. 2006, A\&A, 454L, 21

\bibitem[Higdon et al.(2004)]{smart}
Higdon, S. J. U. et al. 2004, PASP, 116, 975

\bibitem[Imai et al.(2009)]{imai}
Imai, H., He, J.-H., Nakashima, J.-I., Ukita, N., Deguchi, S., Koning, N. 2009, PASJ, 61, 1365

\bibitem[Ivezic et al.(1999)]{dusty}
Ivezic, Z., Nenkova, M.,  Elitzur, M. 1999, User Manual for DUSTY, University of Kentucky Internal Report, accessible at http://www.pa.uky.edu/ ~ moshe/dusty

\bibitem[Kasemann et al.(2006)]{champ}
Kasemann, C., G{\"u}sten, R., Heyminck, S., et al. 2006, in SPIE Conf. Ser., 6275

\bibitem[Kaufman et al.(2006)]{kaufman}
Kaufman, M. J., Wolfire, M. G., Hollenbach, D. J. 2006, ApJ, 644, 283

\bibitem[Klein et al.(2006)]{ffts} 
Klein, B., Philipp, S.~D., Kr{\"a}mer, I., Kasemann, C., G{\"u}sten, R., \& Menten, K.~M.\ 2006, \aap, 454, L29 


\bibitem[Knapp \& Morris(1985)]{knapp}
Knapp, G. M. \& Morris, M. 1985, ApJ, 292, 640

\bibitem[Likkel \& Morris(1988)]{likkel}
Likkel, L. and Morris, M. 1988, ApJ, 329, 914

\bibitem[te Lintel Hekkert et al.(1991)]{telintel91}
te Lintel Hekkert, P., Caswell, J. L., Habing, H. J., Haynes, R. F., Haynes, R. F., Norris, R. P. 1991, A\&ASS, 90, 327

\bibitem[Little-Marenin(1986)]{little}
Little-Marenin, I., R. 1986, ApJ, 307, L15

\bibitem[Ochsenbein et al.(2000)]{vizier}
Ochsenbein F., Bauer P., Marcout J. 2000, A\&AS 143, 221

\bibitem[Oloffson et al.(1993)]{olofsson}
Olofsson, H., Eriksson, K., Gustafsson, B., Carlstrom, U. 1993, ApJS,  87, 267

\bibitem[Oudmaijer et al.(1995)]{oudmaijer}
Oudmaijer, R. D., Waters, L. B. F. M., van der Veen, W. E. C. J.,  Geballe, T. R. 1995, A\&A, 299, 690

\bibitem[Pardo et al.(2001)]{pardo}
Pardo, J. R., Cernicharo, J., \& Serabyn, E. 2001, IEEE Trans. Antennas Propag., 49, 1683

\bibitem[Ramstedt et al.(2008)]{ramstedt}
Ramstedt, S., Sch\"{o}ier, F. L., Olofsson, H.,  Lundgren, A. A. 2008, A\&A, 487, 645

\bibitem[Riscacher et al.(2006)]{flash1}
Risacher, C. et al. 2006, A\&A, 454, L17

\bibitem[Sahai et al.(2007)]{sahai}
Sahai, R., Morris, M., S{\'a}nchez Contreras, C., Claussen, M. 2007, AJ, 134, 2200

\bibitem[Sevenster(2002)]{sevenster}
Sevenster, M. N. 2002, AJ, 123, 2772

\bibitem[Siringo et al.(2009)]{siringo}
Siringo, G. et al.  2009, A\&A, 497, 945

\bibitem[Urquhart et al.(2007)]{urquhart}
 Urquhart, J. S. et al. 2007, A\&A, 474, 891

\bibitem[van der Veen \& Habing(1988)]{vanandhabing}
van der Veen, W. E. C. J. \& Habing, H. J. 1988, A\&A, 194, 125

\bibitem[van der Veen et al.(1989)]{vanderveen}
van der Veen, W. E. C. J., Habing, H.J., Geballe, T. R. 1989, A\&A, 226, 108

\bibitem[Werner et al.(2004)]{spitzer}
Werner, M. W. et al. 2004, ApJS 154, 1
\end{thebibliography}
\end{document}